\documentclass[letterpaper]{article} 
\usepackage{aaai24}  
\usepackage{times}  
\usepackage{helvet}  
\usepackage{courier}  
\usepackage[hyphens]{url}  
\usepackage{graphicx} 
\usepackage{natbib}  
\usepackage{caption} 
\usepackage{algorithm}
\usepackage{algorithmic}
\usepackage{amsfonts}
\usepackage{amsmath}
\usepackage{siunitx}
\usepackage{booktabs}
\usepackage{multirow}
\usepackage{graphicx}
\usepackage{subfigure}
\usepackage{newfloat}
\usepackage{listings}
\DeclareCaptionStyle{ruled}{labelfont=normalfont,labelsep=colon,strut=off} 
\floatstyle{ruled}
\newfloat{listing}{tb}{lst}{}
\floatname{listing}{Listing}
\title{Self-Supervised Disentangled Representation Learning for Robust Target Speech Extraction}
\author {
    Zhaoxi Mu\textsuperscript{\rm 1},
    Xinyu Yang\textsuperscript{\rm 1},
    Sining Sun\textsuperscript{\rm 2},
    Qing Yang\textsuperscript{\rm 2}
}
\affiliations {
    \textsuperscript{\rm 1}Xi'an Jiaotong University\\
    \textsuperscript{\rm 2}Du Xiaoman\\
    wsmzxxh@stu.xjtu.edu.cn, yxyphd@mail.xjtu.edu.cn, \{sunsining,yangqing\}@duxiaoman.com
}

\usepackage{bibentry}

\begin{document}

\maketitle

\begin{abstract}

Speech signals are inherently complex as they encompass both global acoustic characteristics and local semantic information. However, in the task of target speech extraction, certain elements of global and local semantic information in the reference speech, which are irrelevant to speaker identity, can lead to speaker confusion within the speech extraction network. To overcome this challenge, we propose a self-supervised disentangled representation learning method. Our approach tackles this issue through a two-phase process, utilizing a reference speech encoding network and a global information disentanglement network to gradually disentangle the speaker identity information from other irrelevant factors. We exclusively employ the disentangled speaker identity information to guide the speech extraction network. Moreover, we introduce the adaptive modulation Transformer to ensure that the acoustic representation of the mixed signal remains undisturbed by the speaker embeddings. This component incorporates speaker embeddings as conditional information, facilitating natural and efficient guidance for the speech extraction network. Experimental results substantiate the effectiveness of our meticulously crafted approach, showcasing a substantial reduction in the likelihood of speaker confusion.

\end{abstract}

\section{Introduction}

The human auditory system excels in extracting the speech of a target speaker from a complex acoustic environment. Consequently, a longstanding objective of speech-processing research has been to develop machines capable of emulating similar auditory abilities. Target speech extraction (TSE) draws inspiration from human top-down selective auditory attention \cite{mesgarani2012selective,kaya2017modelling}, which employs cues to selectively attend to specific auditory stimuli based on relevance. Discriminative cues utilized in TSE include spatial cues indicating the target speaker's direction \cite{GuCZZXYSZ019}, video recordings of the target speaker's mouth movements \cite{AfourasOCZ20,GaoG21}, and pre-recorded reference speech \cite{WangMWSWHSWJL19,XuRCL20}. Reference speech holds particular value as it provides essential information about the target speaker's voice characteristics and is easily accessible. Accordingly, this paper aims to enhance the performance of monaural TSE methods driven by reference speech.

TSE methods typically comprise two main components: a reference speech encoding network (RSEN) and a speech extraction network (SEN) \cite{ZhaoGS20,XuRCL20}. The RSEN extracts speaker embeddings from the target speaker's reference speech, while the SEN predicts the target speaker's speech within the mixed speech guided by these embeddings. However, the performance of TSE methods often exhibits a long-tail distribution \cite{ZhaoYGZZ22}, indicating that the extracted speech may suffer from the speaker confusion (SC) problem \cite{abs-2202-00733}, also known as target confusion \cite{ZhaoYGZZ22}. The SC problem arises when the RSEN extracts ambiguous speaker embeddings that provide misleading guidance to the SEN. This confusion can cause the SEN to focus on the wrong speaker, leading to inaccurate extraction outcomes.

\begin{figure}[tb]
    \centering
    \includegraphics[width=0.33\textwidth]{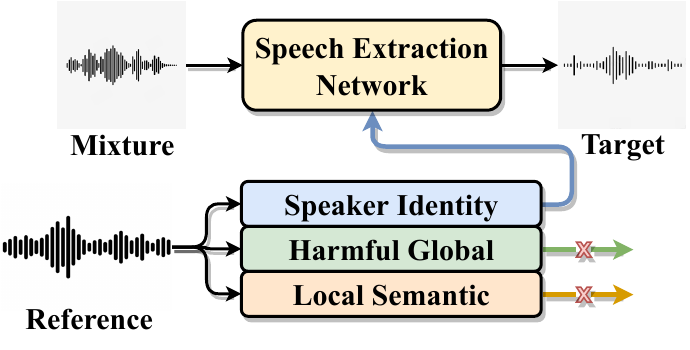}
    \caption{Schematic depicting information disentanglement of reference speech.}
    \label{fig1}
\end{figure}

The SC problem primarily stems from embedding bias \cite{ZhaoYGZZ22}. This bias arises when speaker embeddings extracted from the RSEN fail to accurately represent speaker cues. While these embeddings typically contain discriminative acoustic information associated with the target speaker, they can also be entangled with irrelevant interference information. As speech signals inherently carry both static global acoustic information and dynamic local semantic information, we argue that the reference speech can be disentangled into three distinct components: local information, helpful global information, and harmful global information, as illustrated in Figure \ref{fig1}. Local information pertains to semantic information, while helpful global information refers to speaker identity information. Harmful global information encompasses paralinguistic variables such as emotion, speaking rate, prosody and intonation, which can vary even for the same speaker \cite{DengMSZZSW21,ZhaoYGZZ22}. If semantic information leaks into the embeddings, the SEN may prioritize the source that aligns with the reference speech's semantic information. Likewise, if harmful global information leaks into the embeddings, the SEN may pay more attention to the source that aligns with the reference speech's emotion, speaking rate, prosody, and intonation rather than the source that aligns with the speaker identity of the reference speech. Both of these scenarios can lead to the SC problem.

Previous studies often employ pre-trained speaker recognition networks to extract reference embeddings \cite{ZhangHZ20a,abs-2303-05023}. However, this approach may yield suboptimal embeddings for TSE due to pattern mismatch. Alternatively, the RSEN can be trained using a multi-class cross-entropy loss with speaker identity labels and jointly optimized with the SEN \cite{ZhaoGS20,XuRCL20}. However, a limitation of this method is its reliance on speaker identity labels in the training data, which may not be available in real-world scenarios.

To address the challenges above, we propose a novel two-phase \textbf{s}elf-supervised \textbf{d}isentangled \textbf{r}epresentation learning (DRL) method for robust \textbf{t}arget \textbf{s}peech \textbf{e}xtraction, called SDR-TSE. Our approach involves explicitly disentangling the reference speech's semantic and global information using the RSEN, followed by the implicit disentanglement of the speaker identity information within the global information using the global information disentanglement network (GIDN). By only utilizing the disentangled information for guidance, our TSE pipeline avoids any leakage of harmful information from the reference speech. Moreover, the RSEN and GIDN are trained in a self-supervised manner, negating the dependence on speaker identity labels and enhancing the applicability across diverse real-world scenarios.

Previous methods for integrating speaker embeddings and acoustic representations in TSE typically relied on simplistic summation and concatenation methods \cite{GeXWCD020,DengMSZZSW21}. However, these methods are susceptible to information overload, as speaker embeddings can overwhelm the acoustic representation information. To overcome this limitation, we introduce a natural fusion method by replacing the layer normalization in the Transformer with adaptive modulation layer normalization (AMLN). AMLN integrates speaker embeddings as conditional information to enhance the SEN's perception capability for the target speaker without interfering with the acoustic representation.

In summary, this paper makes several notable contributions: (\romannumeral1) We propose a novel two-phase self-supervised DRL policy to effectively tackle the issue of speaker confusion in TSE. (\romannumeral2) We propose an approach for incorporating speaker identity information into the SEN naturally and efficiently. (\romannumeral3) We conduct comprehensive experiments to validate the significance of information disentanglement, and our method defines new state-of-the-art performance.

\section{Related Work}

\textbf{Speech Separation.} Speech separation (SS) refers to the process of isolating individual speech components from mixed speech signals \cite{WangC18a}. Early advancements in SS primarily focused on techniques in the time-frequency domain \cite{HersheyCRW16,YuKT017,KolbaekYTJ17}. To circumvent the explicit phase estimation problem, \citeauthor{LuoM19} \shortcite{LuoM19} proposed Conv-TasNet, a time-domain SS model that employs a CNN to extract speech features. Additionally, to handle the separation of long speech sequences while reducing computational complexity, \citeauthor{LuoCY20} \shortcite{LuoCY20} proposed the DPRNN, consisting of three components: segmentation, chunk processing, and overlap-add. Recently, \citeauthor{SubakanRCBZ21} \shortcite{SubakanRCBZ21} proposed the Sepformer, which replaces the RNN in DPRNN with a Transformer architecture.

\textbf{Target Speech Extraction.} TSE tackles the challenges of unknown speaker numbers and speaker permutations, which are not fully resolved in SS methods, by leveraging the target speaker's reference speech. Previous research has investigated the SC problem in TSE. For instance, \citeauthor{ZhaoYGZZ22} \shortcite{ZhaoYGZZ22} proposed a two-stage solution. In the training stage, they integrated metric learning methods to enhance the discriminability of the embeddings extracted by the RSEN. In the inference stage, they employed a post-filtering strategy to rectify erroneous results. However, this method is limited to scenarios involving two speakers in mixed speech. More recently, \citeauthor{abs-2303-05023} \shortcite{abs-2303-05023} introduced two novel loss functions to optimize performance metrics by defining the reconstruction quality at the chunk level. These loss functions make use of metric-correlated distribution information, enabling the SEN to focus on the chunks where SC occurs.

\textbf{Disentangled Representation Learning.} DRL aims to separate unique, independent, and informative factors present in the data. Disentangled latent variables exhibit sensitivity to changes in a single underlying factor while insensitive to other factors, thus ensuring statistical independence \cite{BengioCV13,abs-2211-11695}. DRL has found widespread applications in speech synthesis, conversion, and enhancement to enhance model interpretability and controllability. For instance, \citeauthor{ChoiLKLHL21}  \shortcite{ChoiLKLHL21} and \citeauthor{QianZGNLCHC22} \shortcite{QianZGNLCHC22} utilized information perturbation as a data enhancement technique to learn speaker-independent feature representations and disentangle the speaker information from other information in speech. However, these approaches rely on pre-trained speech feature extractors. \citeauthor{HouXC021} \shortcite{HouXC021} addressed the issue of noise type mismatch in speech enhancement by employing a noise type classifier with a gradient reversal layer (GRL) as the disentangler to learn noise-agnostic feature representations. Similarly, \citeauthor{NekvindaD20} \shortcite{NekvindaD20} and \citeauthor{abs-2305-19522} \shortcite{abs-2305-19522} also employed GRL to disentangle speaker identity information from speech signals. However, these methods rely on noise type or speaker identity labels for training, while our proposed DRL method is self-supervised.

\section{Methodology}

\subsection{Notations and Problem Formulation}

The mixed speech signal $y \in \mathbb{R}^{T}$ of length $T$, comprising multiple speakers and noise interference, can be expressed as the sum of the target speech $u \in \mathbb{R}^{T}$ and other interfering components $v \in \mathbb{R}^{T}$, 
\begin{equation}
y=u+v
\end{equation}
The objective is to separate the target speech $u$ from other interfering components in $y$ using the target speaker's cues,
\begin{equation}
\hat{u}=\mathcal{F}(y,z_s;\theta_{\mathcal{F}})
\end{equation}
where $\hat{u}$ represents the target speech predicted by the SEN $\mathcal{F}$, and $\theta_{\mathcal{F}}$ denotes the parameters of $\mathcal{F}$. The target speaker embedding $z_s$ is utilized to guide $\mathcal{F}$ and is extracted by encoding a reference speech $x$ from the same speaker as $u$.

\subsection{Model Overview}

\begin{figure*}[tb]
    \centering
    \includegraphics[width=0.95\textwidth]{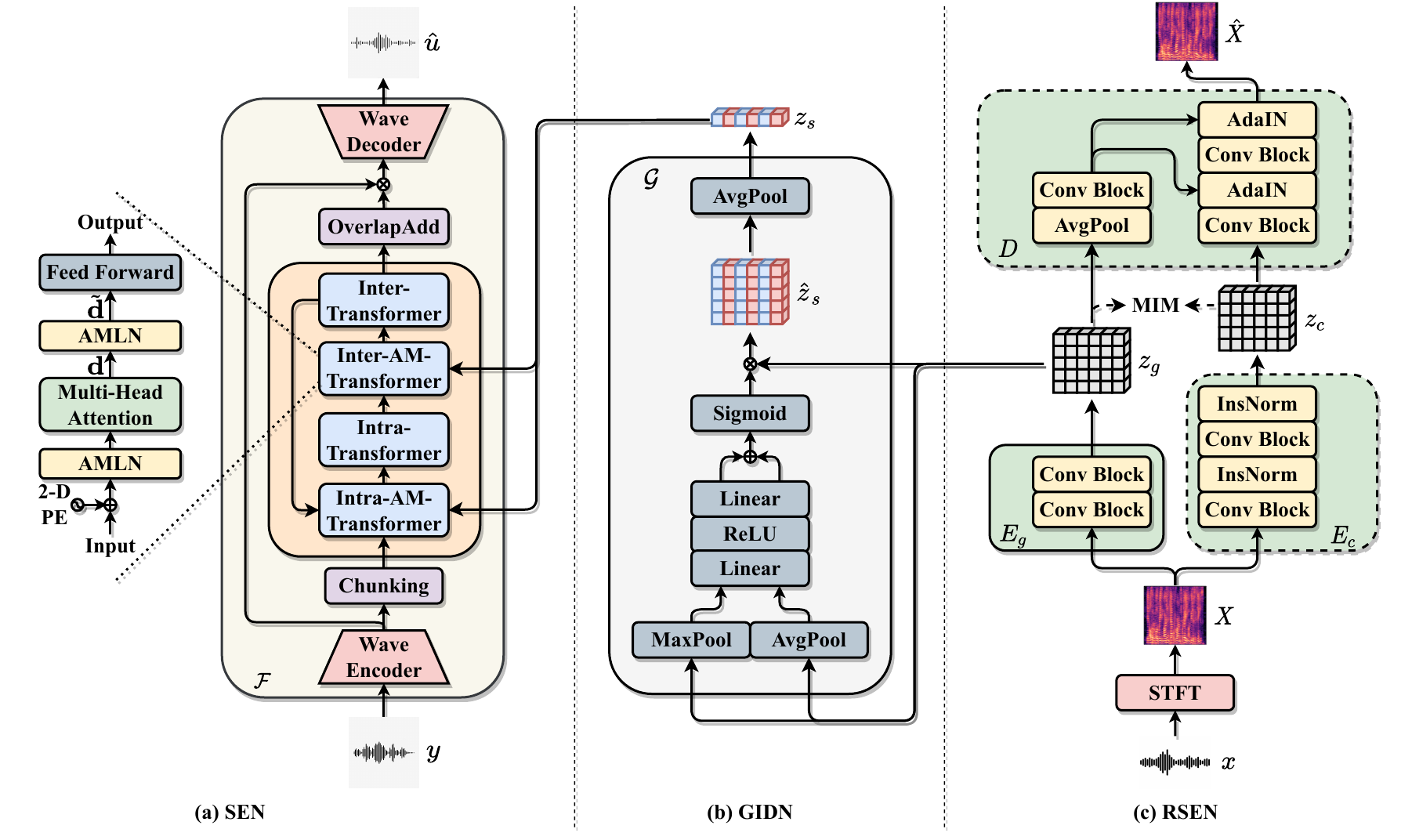}
    \caption{The architecture of the SDR-TSE. (a), (b) and (c) depict the speech extraction network, global information disentanglement network and reference speech encoding network. The semantic information encoder $E_c$ and spectrogram decoder $D$ within the dashed box are utilized solely for training purposes to facilitate disentanglement and discarded during inference. MIM refers to mutual information minimization. The red channels in the feature map of the GIDN indicate activated channels containing speaker identity information, while blue channels represent suppressed channels containing harmful information.}
    \label{fig2}
\end{figure*}

Figure \ref{fig2} illustrates the architecture of our proposed model, which consists of the SEN, RSEN and GIDN. The SEN is built upon the state-of-the-art (SOTA) SS model Sepformer \cite{SubakanRCBZ21}, which follows an encoder-masker-decoder framework, as depicted in Figure \ref{fig2}(a). The waveform encoder of the SEN consists of a 1-D convolution that encodes $y$ into a time-domain feature representation. This feature representation is subsequently segmented into overlapping chunks, concatenated and fed into stacked Transformer blocks to alternate between local and global modelling. The resulting output is then transformed back into sequential permutation using the overlap-add method, enabling the prediction of the target speaker mask. The mask is multiplied with the feature representation of $y$ encoded by the waveform encoder to derive the feature representation of the target speaker. Ultimately, this feature representation is fed into the waveform decoder, composed of a 1-D transposed convolution, to generate the target speaker's speech.

The RSEN is a vital component of the SDR-TSE, comprising three main components: a global information encoder $E_g$, a semantic information encoder $E_c$, and a spectrogram decoder $D$, as displayed in Figure \ref{fig2}(c). $E_g$ and $E_c$ encode the reference speech into global and semantic spaces, respectively. $D$ reconstructs the spectrogram of the reference speech using the global and semantic information representations. The global information representation produced by $E_g$ is then fed into the GIDN to extract the speaker embedding, as depicted in Figure \ref{fig2}(b). The speaker embedding serves as a guidance signal in the Intra- and Inter-AM-Transformer blocks of the SEN. These components will be described in more detail in the subsequent sections.

\subsection{Reference Speech Encoding Network}

Let $x \in \mathbb{R}^{T_x}$ represent the reference speech, and $X \in \mathbb{R}^{F_X \times T_X}$ denote the spectrogram of $x$ generated via Short-Time Fourier Transform (STFT). The global and semantic latent representations of $X$ are denoted as $z_g \in \mathbb{R}^{d_g \times T_g}$ and $z_c \in \mathbb{R}^{d_c \times T_c}$, respectively. To disentangle these two latent representations, we assume they are probabilistically independent. We define the joint latent representation as $z=[z_g,z_c]$ and factorize the prior distribution $p(z)$ and posterior distribution $p(z\mid X)$ of $z_g$ and $z_c$ by following the independence assumption:
\begin{equation}
p(z)=p(z_g)p(z_c)
\end{equation}
\begin{equation}
p(z\mid X)=p(z_g\mid X)p(z_c\mid X)
\end{equation}
$E_g$ and $E_c$ estimate the posterior distributions $p(z_g\mid X)$ and $p(z_c\mid X)$ as $q(z_g\mid X)$ and $q(z_c\mid X)$. The variational autoencoder (VAE), known for its capability to disentangle information and model the semantic information of speech \cite{abs-2211-11695}, is employed to construct the RSEN. To learn meaningful semantic information representation, we assume that the prior $p(z_c)$ of semantic information follows a standard normal distribution.

$E_g$ is responsible for encoding $X$ into a global feature space, producing the global information representation $z_g$. This process can be expressed as:
\begin{equation}
q(z_g\mid X)=E_g(X;\theta_{E_g})
\end{equation}
$\theta_{E_g}$ represents the parameters of $E_g$. $E_g$ aims to capture all global features of the reference speech. To enable $E_c$ to encode $X$ into a semantic space and acquire a meaningful semantic representation $z_c$, we assume $q(z_c\mid X)$ follows a conditionally independent Gaussian distribution with unit variance to reduce complexity, motivated by \citeauthor{LiuBK17} \shortcite{LiuBK17} and \citeauthor{ChouL19} \shortcite{ChouL19}. Formally, we express this as follows:
\begin{equation}
q(z_c \mid X)=\mathcal{N}(E_c(X;\theta_{E_c}),I)
\end{equation}
where $\theta_{E_c}$ represents the parameters of $E_c$. The role of $D$ is to reconstruct the spectrogram $\hat{X}$ of the reference speech $x$ using the encoded representations $z_g$ and $z_c$. The reconstruction process can be formulated as follows:
\begin{equation}
\hat{X}=D(z_g,z_c;\theta_{D})
\end{equation}
$\theta_{D}$ represents the parameters of $D$. The reconstruction step allows $E_g$ and $E_c$ to effectively encode global and semantic information of the reference speech. To achieve this, we apply average pooling on the temporal dimension of $z_g$ and leverage the technique of adaptive instance normalization (AdaIN) \cite{ChouL19} to conditionally reconstruct $\hat{X}$ using the pooled embedding vector.

The RSEN is optimized by maximizing the Evidence Lower Bound (ELBO):
\begin{equation}
\begin{aligned}
\label{eq8}
\max_{\theta_{E_c},\theta_{E_g},\theta_{D}} \textbf{ELBO} = \mathbb{E}_{q(z_c\mid X)}\mathbb{E}_{q(z_g\mid X)}(\log p(X\mid z_c,z_g)) \\ 
- D_{KL} \left( q(z_c\mid X) \parallel p(z_c) \right)
\end{aligned}
\end{equation}
$D_{KL}$ represents the Kullback-Leibler (KL) divergence. Equivalently, Eq.(\ref{eq8}) can be optimized by minimizing the reconstruction loss $\mathcal{L}_{\text{REC}}$ and the KL divergence loss $\mathcal{L}_{\text{KL}}$:
\begin{equation}
\mathcal{L}_{\text{REC}}= \| \hat{X}-X  \|_1
\end{equation}
\begin{equation}
\mathcal{L}_{\text{KL}}= \| z_c  \|^2_2
\end{equation}
$\| \cdot \|_1$ and $ \| \cdot \|_2$ represent the $L^1$ norm and $L^2$ norm. To alleviate computational complexity, we calculate $\mathcal{L}_{\text{REC}}$ in the time-frequency domain of the speech signal. $\mathcal{L}_{\text{KL}}$ encourages the posterior $q(z_c\mid X)$ to align with the prior $p(z_c) = \mathcal{N}(z_c \mid 0,I)$, where $I$ denotes the identity matrix. Notably, the training process is self-supervised and does not rely on speaker identity labels. $E_c$ and $D$ are exclusively utilized during training and discarded during inference.

\textbf{Disentangled Representation Learning.} In the RSEN, we employ various techniques to disentangle the latent variables $z_g$ and $z_c$. Firstly, we introduce instance normalization (IN) in $E_c$ as an information bottleneck to filter out global information while preserving the semantic information, motivated by \citeauthor{ChouL19} \shortcite{ChouL19} and \citeauthor{ChenWWL21} \shortcite{ChenWWL21}. Furthermore, during training, we minimize the mutual information (MI) between $z_g$ and $z_c$ to prevent any mutual leakage between the semantic and global information. Incorporating MI as a regularization term can enhance the disentanglement capability of the VAE and constrain the dependencies between $z_g$ and $z_c$. Specifically, in the previous section, we assumed that the random variables $z_c$ and $z_g$ are mutually independent. To achieve this, we minimize the KL divergence between their joint distribution and the product of their marginal distributions, which can be expressed as:
\begin{equation}
\label{eq11}
\min_{\theta_{E_c},\theta_{E_g}} \mathcal{I}(z_c,z_g)=\mathbb{E}_{p(z_c,z_g)} \left[ \log \frac{p(z_c,z_g)}{p(z_c)p(z_g)} \right] 
\end{equation}
$\mathcal{I}(z_c,z_g)$ represents the MI between $z_c$ and $z_g$. Eq.(\ref{eq11}) can also function as a regulariser of Eq.(\ref{eq8}). We employ the vCLUB method \cite{ChengHDLGC20} to estimate an upper bound on the MI, which is given by:
\begin{equation}
\begin{aligned}
\mathcal{I}_{\text{vCLUB}}(z_c,z_g)=\mathbb{E}_{p(z_c,z_g)} \left[ \log q(z_c\mid z_g) \right] \\
- \mathbb{E}_{p(z_c)}\mathbb{E}_{p(z_g)}\left[ \log q(z_c\mid z_g)  \right]
\end{aligned}
\end{equation}
$q(z_c\mid z_g)$ serves as a variational approximation to the true posterior distribution $p(z_c\mid z_g)$, and is parameterized by a neural network $\mathcal{V}$. An unbiased estimator of the vCLUB between $z_c$ and $z_g$ can be given by:
\begin{equation}
\begin{aligned}
\hat{\mathcal{I}}_{\text{vCLUB}}(z_c,z_g)= \frac{1}{N} \sum_{i=1}^{N} \left[ \log q(z_{c}^{(i)}\mid z_{g}^{(i)}) \right. \\ 
-  \frac{1}{N} \sum_{j=1}^N \log q(z_{c}^{(j)}\mid z_{g}^{(i)})  \left.\right]
\end{aligned}
\end{equation}
$N$ denotes the number of samples. $\hat{\mathcal{I}}_{\text{vCLUB}}(z_c,z_g)$ provides a reliable upper bound on MI with a well-performing variational approximation. To improve the accuracy of the vCLUB MI estimator, we train the variational approximation network $\mathcal{V}$ to maximize the log-likelihood:
\begin{equation}
\label{LL}
\mathcal{L}_{\text{LL}}= \frac{1}{N} \sum_{i=1}^{N} \log q(z_{c}^{(i)}\mid z_{g}^{(i)})
\end{equation}
$\mathcal{V}$ and RSEN are alternately optimized during training.

\subsection{Global Information Disentanglement Network}

We consider that each channel of the output $z_g$ from $E_g$ encompasses distinct types of information \cite{abs-2303-03737,abs-2303-03732}. Some channels may contain valuable speaker identity information, while others may contain harmful global information. To disentangle the useful speaker identity information, the GIDN $\mathcal{G}$ is utilized to refine $z_g$ further. The GIDN extracts the speaker embedding $z_s \in \mathbb{R}^{d_s}$, exclusively containing speaker identity information for guiding the SEN. Specifically, we employ channel attention \cite{WooPLK18} to adjust the importance of individual channels within $z_g$ to focus on channels that contribute to the TSE while suppressing irrelevant channels. The weight assigned to each channel of $z_g$ can be calculated as:
\begin{equation}
\phi(z_g)=\sigma(W_1f(W_0(z_g^{\text{avg}}))+W_1f(W_0(z_g^{\text{max}})))
\end{equation}
$z_g^{\text{avg}}$ and $z_g^{\text{max}}$ are derived from $z_g$ by performing average and max pooling operations. Afterwards, they are fed into two linear layers that share weights $W_0$ and $W_1$. The activation function $f$ used is ReLU, and $\sigma$ denotes the sigmoid function. The resulting output $\hat{z}_s$ of the channel attention is:
\begin{equation}
\hat{z}_s=\phi(z_g)\otimes z_g
\end{equation}
$\otimes$ denotes element-wise multiplication. We apply average pooling along the temporal dimension of $\hat{z}_s$ to obtain the time-independent speaker embedding $z_s$ to guide the SEN.

To ensure that $\mathcal{G}$ selectively activates the channels in $z_g$ that contain speaker identity information while suppressing those with harmful global information, we incorporate the concept of contrastive learning for training $\mathcal{G}$. The objective is to guarantee that the reference speech $x$ belongs to the same speaker as $\hat{u}$ extracted by the SEN, while the interference signal $\hat{v}$, computed as $y-\hat{u}$, belongs to other speakers. Thus, the speaker embedding $z_s^{(x)}$ of the reference speech $x$, extracted by $\mathcal{G}$, should exhibit similarities to the embedding $z_s^{(\hat{u})}$ of $\hat{u}$ while being distinct from the embedding $z_s^{(\hat{v})}$ of $\hat{v}$. To enforce this constraint, we employ a similarity discriminative loss $\mathcal{L}_{\text{SIM}}$ to train $\mathcal{G}$, defined as follows:
\begin{equation}
\mathcal{L}_{\text{SIM}}=\langle z_s^{(x)},z_s^{(\hat{u})}\rangle -\langle z_s^{(x)},z_s^{(\hat{v})}\rangle
\end{equation}
$\langle \cdot,\cdot \rangle$ is the dot product operator, and cosine similarity is employed to quantify the similarity between two embeddings. Notably, when back-propagating $\mathcal{L}_{\text{SIM}}$, the parameters of the SEN are held constant, while the parameters of the GIDN and $E_g$ are optimized.

\subsection{Speech Extraction Network}

To effectively utilize $z_s$ to guide the SEN $\mathcal{F}$, we propose the Adaptive Modulation Transformer (AM-Transformer) as a replacement for the Transformer module in Sepformer, motivated by \citeauthor{MinLYH21} \shortcite{MinLYH21} and \citeauthor{Wu00HZSQL22} \shortcite{Wu00HZSQL22}. The AM-Transformer is capable of naturally incorporating speaker identity information as a condition. As depicted in Figure \ref{fig2}(a), we substitute the layer normalization \cite{BaKH16} in Transformer with adaptive modulation layer normalization (AMLN). In contrast to the fixed gain and bias in layer normalization, we leverage $z_s \in \mathbb{R}^{d_s}$ as a condition to predict the gain and bias of the input acoustic representation. Specifically, given the input acoustic representation $\mathbf{d} \in \mathbb{R}^{H\times T_d}$ for the AMLN, we calculate its mean $\mu \in \mathbb{R}^{T_d}$ and standard deviation $\sigma \in \mathbb{R}^{T_d}$. The normalized vector $\mathbf{h} \in \mathbb{R}^{H\times T_d}$ of $\mathbf{d}$ is defined as:
\begin{equation}
\mathbf{h}=\frac{\mathbf{d}-\mu}{\sigma}
\end{equation}
The output of AMLN, denoted as $\tilde{\mathbf{d}} \in \mathbb{R}^{H\times T_d}$, is given by:
\begin{equation}
\tilde{\mathbf{d}}=\gamma(z_s)\cdot \mathbf{h}+\beta(z_s)
\end{equation}
$\gamma(z_s) \in \mathbb{R}^{H}$ and $\beta(z_s) \in \mathbb{R}^{H}$ represent two affine transformations of $z_s$, implemented by two fully connected layers. They adaptively scale and shift $\mathbf{h}$ based on the condition $z_s$. Additionally, we employ 2-D position encoding (PE) \cite{raisi20202d,abs-2306-14170} instead of the original PE. This modification enables more effective utilization of intra- and inter-chunk positional information. The entire model is optimized using the SI-SNR loss $\mathcal{L}_{\text{SI-SNR}}$ \cite{RouxWEH19}. The training process is shown in Algorithm \ref{alg}. 

\begin{algorithm}[tb]
\caption{SDR-TSE Optimization}
\label{alg}
\textbf{Require}: The training data $D^{\star}$ containing mixed-target-reference speech triplets ($y$, $u$, $x$).
\begin{algorithmic}[1] 
\STATE Initialize the entire system randomly.
\WHILE{not converged}
\STATE Sample $\{(y_i, u_i, x_i)\}_{i=1}^{N}$ from $D^{\star}$.
\STATE \textbf{\emph{Forward-Propagation}}
\STATE Reconstruct the spectrogram $\{\hat{X}_i\}_{i=1}^{N}$ of $\{x_i\}_{i=1}^{N}$ and predict the target speech $\{\hat{u}_i\}_{i=1}^{N}$.
\STATE \textbf{\emph{Back-Propagation}}
\STATE Update $\theta_{\mathcal{V}}$ by maximizing $\mathcal{L}_{\text{LL}}$.
\STATE Update $\theta_{E_g}$, $\theta_{E_c}$, $\theta_{D}$, $\theta_{\mathcal{G}}$ and $\theta_{\mathcal{F}}$ by minimizing $\mathcal{L}_{\text{KL}}$, $\mathcal{L}_{\text{REC}}$, $\mathcal{I}_{\text{vCLUB}}$ and $\mathcal{L}_{\text{SI-SNR}}$.
\STATE Update $\theta_{\mathcal{G}}$ and $\theta_{E_g}$ by minimizing $\mathcal{L}_{\text{SIM}}$.
\ENDWHILE
\end{algorithmic}
\end{algorithm}

\section{Experiments}

\subsection{Datasets and Implementation Details}

The TSE model is trained and evaluated using the widely-used two-speaker mixed dataset WSJ0-2mix \cite{HersheyCRW16} and its derivative dataset WSJ0-2mix-extr \cite{XuRCL20}. WSJ0-2mix-extr is utilized for all ablation studies.

In the SEN, each AM-Transformer layer is accompanied by $7$ Transformer layers. The Intra-Inter iteration is repeated twice. For the RSEN, we employ convolutional blocks stacked by 1-D convolutional layers to construct $E_g$, $E_c$, and $D$. The reference speech $x$ is processed using STFT with a window length of $512$, a hop length of $128$, and an STFT window size of $512$. The resulting spectrogram is then converted to a magnitude spectrogram so that $F_x$ is $257$. The dimensions $d_g$, $d_c$, $d_s$, and $H$ are all set to $256$. The variational approximation network $\mathcal{V}$ is implemented using two four-layer fully connected networks to predict the mean and variance of the posterior distribution, respectively. The model encompasses a total of $45$M parameters. The weights of $\mathcal{L}_{\text{SI-SNR}}$, $\mathcal{L}_{\text{REC}}$, $\mathcal{L}_{\text{KL}}$, $\mathcal{I}_{\text{vCLUB}}$, $\mathcal{L}_{\text{LL}}$ and $\mathcal{L}_{\text{SIM}}$ are set to $1$, $\num{e-3}$, $\num{e-4}$, $\num{e-4}$, $\num{e-3}$ and $\num{e-3}$, respectively, determined through a grid search.

\subsection{Evaluation Metrics}

To facilitate comparison with other methods, the performance of TSE is evaluated by SI-SNRi and SDRi \cite{VincentGF06}, while speech quality is assessed by PESQ \cite{RixBHH01}. During our experiment, we observed that the occurrence probability of SC (SI-SNRi is negative) is not high for the entire extracted speech, but SC often occurs in specific speech chunks. Therefore, it is more appropriate to utilize the negative SI-SNRi ratio (NSR) for segmented speech rather than the entire speech \cite{ZhangHZ20a,ZhaoYGZZ22} to measure the probability of SC occurrence. To quantify this, we employ the chunk-wise SC measure metric $r_{\text{scr}}$ \cite{abs-2303-05023}, which is defined as the ratio of the number of speech chunks with SC to the total number of active speech chunks:
\begin{equation}
M=\left\lceil \frac{T-L}{O}+1 \right\rceil
\end{equation}
\begin{equation}
S(k) = s(\hat{u}(k),u(k))-s(y(k)),u(k))
\end{equation}
\begin{equation}
N_{\text{sc}}=\sum_{k=1}^{M}\mathbb{I}(S(k)<0)
\end{equation}
\begin{equation}
N_{\text{vaild}}=\sum_{k=1}^{M}\mathbb{I}(\mathrm{E}(u(k))>\eta)\cdot \mathbb{I}(\mathrm{E}(\hat{u}(k))>\eta)
\end{equation}
\begin{equation}
r_{\text{scr}}=\frac{N_{\text{sc}}}{N_{\text{vaild}}}\times 100\%
\end{equation}
To calculate $r_{\text{scr}}$, the ground truth target speech $u$, predicted target speech $\hat{u}$, and mixture $y$ need to be segmented into chunks. The total number of chunks $M$ is determined by the speech length $T$, chunk length $L$, and hop length $O$. $\lceil \cdot \rceil$ denotes the ceiling function. For each chunk $k=1,2,\dots, M$, we calculate the chunk-level SI-SNR improvement $S(k)$, where $s$ represents the SI-SNR. If $S(k)<0$, it is considered that SC occurred in the $k$th chunk. $\mathbb{I}$ denotes the indicator function. $\mathrm{E}$ represents the energy of the speech, and $\eta$ is the energy-related threshold. $L$, $O$, and $\eta$ are set to $250$ ms, $125$ ms, and $5\%$ of the maximum energy, respectively.

\subsection{Investigation of Disentangled Representations}

\begin{figure*}[tb]
    \centering
    \includegraphics[width=0.64\textwidth]{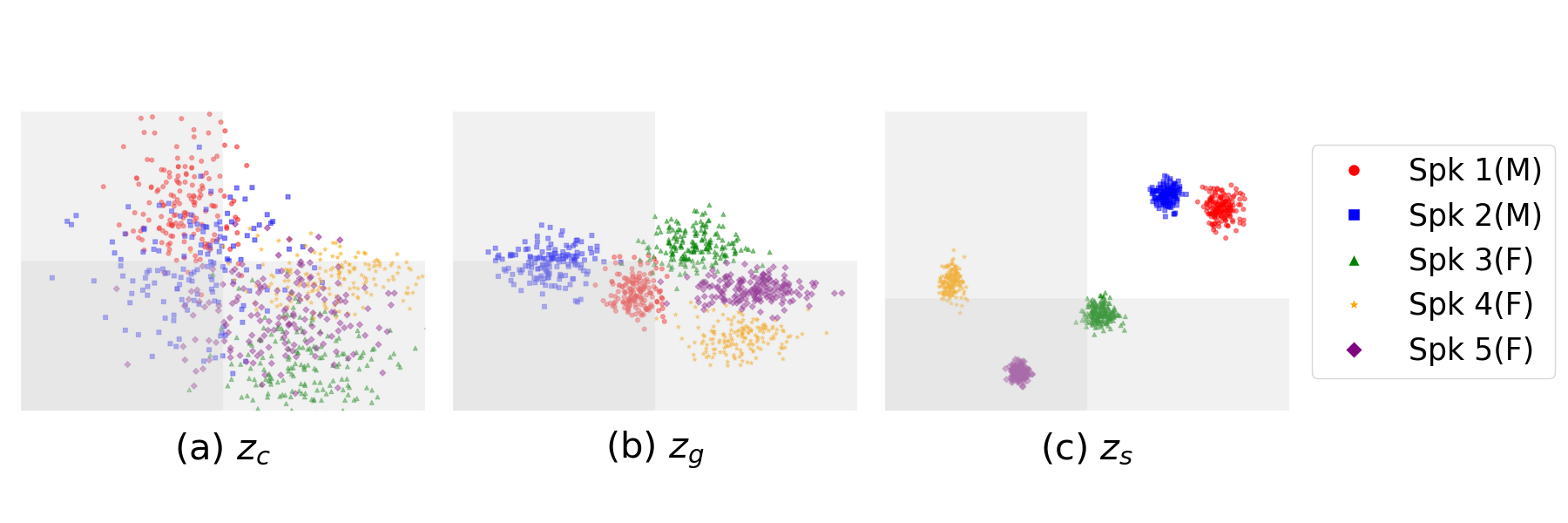}
    \caption{2-D visualization of the spatial distribution of $z_c$, $z_g$, and $z_s$ for the reference speech of five different speakers on the WSJ0-2mix-extr test set. Speakers are labelled as M (male) and F (female).}    
    \label{fig6}    
\end{figure*}

To intuitively illustrate the information captured in the embedding vectors of reference speech at each phase, we employ t-SNE to visualize the spatial distribution of $z_c$, $z_g$, and $z_s$, as depicted in Figure \ref{fig6}(a), (b), and (c) respectively. It is apparent that $z_c$ contains substantial overlapping information across different speakers, indicating its representation of speaker-independent semantic information. Although $z_g$ exhibits speaker clustering to some extent, there is still a small overlap among different speakers, implying that the global information of speech represented by $z_g$ may share similar characteristics across various speakers. In contrast, $z_s$ exhibits distinct speaker clustering, with well-defined boundaries separating each speaker and notable distribution differences between male and female speakers. These results suggest that $z_s$ effectively represents speaker identity.

To further demonstrate the significance of information disentanglement, we employ each disentangled information individually to guide the same speech extraction network, as depicted in Table \ref{tab1}. The AMLN method is utilized to modulate the SEN with $z_g$ and $z_s$. For $z_g$, average pooling is applied along its temporal dimension, resulting in an embedding vector that serves as a condition. Cross-attention mechanism \cite{abs-2303-05023,abs-2306-14170} is employed to fuse $z_c$ with the input of both the Intra- and Inter-Transformer to avoid the disruption of content information with dynamic local features caused by global pooling.

\begin{table}[tb]
    \centering
    \small
    \setlength\tabcolsep{2pt} 
    \begin{tabular}{c|ccc|ccc|c}
        \toprule
        ID  & CI ($z_c$) & GI ($z_g$) & SI ($z_s$) & SI-SNRi$\uparrow$ & SDRi$\uparrow$ & PESQ $\uparrow$ & $r_{\text{scr}}$ $\downarrow$\\ 
        \midrule
        1 & $\times$ & $\times$ & $\checkmark$ & $\textbf{19.9}$ & $\textbf{20.2}$ & $\textbf{3.85}$ & $\textbf{6.89}$ \\
        2 & $\times$ & $\checkmark$ & $\times$ & $19.3$  & $19.6$ & $3.81$ & $7.22$ \\
        3 & $\checkmark$ & $\times$ & $\times$ & $15.6$  &  $15.9$ & $3.50$ & $12.14$ \\
        4 & $\checkmark$ & $\checkmark$ & $\times$ & $17.5$   & $17.8$ & $3.61$ & $9.06$ \\
        5 & $\checkmark$ & $\times$ & $\checkmark$ & $17.9$  & $18.3$ & $3.63$ & $8.35$ \\
        \bottomrule
    \end{tabular}
    \caption{Experimental results involve using various information representations and their combination to guide the SEN on the WSJ0-2mix-extr dataset. CI, GI, and SI represent semantic, global, and speaker identity information.}
    \label{tab1}
\end{table}

The findings in Table \ref{tab1} demonstrate that utilizing only the disentangled $z_s$ to guide the SEN yields the most favourable outcomes with the lowest SC ratio (ID-1). When employing $z_g$ as the guidance without the GIDN (ID-2), all performance metrics deteriorate, confirming the channel attention's capability to suppress harmful global information. Using solely $z_c$ as the guidance (ID-3) yields the poorest results, particularly with $r_{\text{scr}}$ increasing to $12.14$, indicating a severe SC problem. This signifies that the semantic information in the reference speech offers little guidance in aiding TSE. When combining $z_c$ with either $z_g$ or $z_s$ (ID-4 and ID-5), the performance metrics achieve suboptimal levels compared to ID-2 and ID-1, further affirming the necessity of information disentanglement. The results in Table \ref{tab1} align with the distribution of each embedding illustrated in Figure \ref{fig6}.

\subsection{Ablation Study}

To assess the effectiveness of each module and loss function in our proposed method, we conducted ablation experiments on the IN layer of $E_c$, MI minimization, $\mathcal{L}_{\text{SIM}}$, $\mathcal{L}_{\text{REC}}$, $\mathcal{L}_{\text{KL}}$, and 2-D PE. Each ablation experiment was conducted independently while keeping all other components unchanged. The results are summarized in Table \ref{tab2}. It is evident that all the components contribute to improving the overall performance. The IN layer acts as an information bottleneck for $E_c$ to help filter out global information. MI minimization filters out semantic and global information from the output of $E_g$ and $E_c$, respectively. $\mathcal{L}_{\text{SIM}}$ filters out harmful global information from the speaker identity information. These three components complement each other, and all contribute positively to information disentanglement. The inclusion of $\mathcal{L}_{\text{REC}}$ and $\mathcal{L}_{\text{KL}}$ ensures that $E_g$ and $E_c$ learn meaningful global and semantic features, respectively. The utilization of 2-D PE also enhances the performance of the SEN by facilitating better comprehension of the relative positional relationships at both intra- and inter-chunk time steps.

\begin{table}[tb]
    \centering
    \small
    \setlength\tabcolsep{3pt}
    \begin{tabular}{l|ccc|c}
        \toprule
        Method  & SI-SNRi$\uparrow$ & SDRi$\uparrow$  & PESQ $\uparrow$  & $r_{\text{scr}}$ $\downarrow$\\
        \midrule
        SDR-TSE & $\textbf{19.9}$  & $\textbf{20.2}$ &  $\textbf{3.85}$ & $\textbf{6.89}$   \\
        w/o IN & $18.2$  & $18.4$   &  $3.69$ & $8.55$  \\
        w/o MIM & $17.3$  & $17.6$   &  $3.59$ & $8.60$  \\
        w/o $\mathcal{L}_{\text{SIM}}$  & $18.4$  & $18.7$ &  $3.72$ & $8.36$    \\
        w/o $\mathcal{L}_{\text{REC}}$  & $16.2$  & $16.5$ &  $3.50$ & $9.92$    \\
        w/o $\mathcal{L}_{\text{KL}}$  & $18.8$  & $19.1$ &  $3.78$ & $7.61$    \\
        w/o 2-D PE & $19.4$  & $19.6$   &  $3.81$ & $7.27$  \\
        \bottomrule
    \end{tabular}
    \caption{Results of ablation experiments on the WSJ0-2mix-extr dataset.}
    \label{tab2}
\end{table}

Next, we compared various modulation policies for acoustic representations in SEN and speaker embeddings, as outlined in Table \ref{tab3}. The results indicate that simplistic addition and concatenation methods yield the worst performance. This is because such approaches fail to ensure the preservation of undisturbed acoustic representation information during fusion. Compared to Gated Conv \cite{LiuX22a} and ConSM \cite{abs-2306-16250}, our method possesses the advantage of applying layer normalization to the acoustic representation prior to adaptive modulation. This sequential process enhances the perception of speaker embeddings by preserving their inherent identity information, as we observed that normalizing after modulation could disrupt speaker identity information.

\begin{table}[tb]
    \centering
    \small
    \setlength\tabcolsep{3pt}    
    \begin{tabular}{l|ccc|c}
        \toprule
        Modulation Policy  & SI-SNRi$\uparrow$ & SDRi$\uparrow$ & PESQ $\uparrow$ & $r_{\text{scr}}$ $\downarrow$\\
        \midrule
        Summation          & $18.7$ & $18.9$ & $3.72$ &  $8.23$  \\
        Concatenation       & $19.0$ & $19.3$ & $3.74$ & $8.07$ \\
        Gated Conv \shortcite{LiuX22a} & $19.3$ & $19.5$  & $3.80$ &  $7.63$ \\
        ConSM \shortcite{abs-2306-16250}       & $19.2$ & $19.5$  & $3.77$ &  $7.88$  \\
        AMLN (Ours) & $\textbf{19.9}$   & $\textbf{20.2}$ & $\textbf{3.85}$ & $\textbf{6.89}$  \\
        \bottomrule
    \end{tabular}
    \caption{Experimental results of employing various modulation policies on the WSJ0-2mix-extr dataset.}
    \label{tab3}
\end{table}

\subsection{Comparison with the State-of-the-Art}

We compared our method with SOTA TSE methods on the WSJ0-2mix-extr and WSJ0-2mix datasets, and the results are summarized in Table \ref{tab4} and Table \ref{tab5}. Our method outperforms other methods across all metrics. This performance improvement can be attributed to two key factors. Firstly, we employ the SOTA SS framework Sepformer \cite{SubakanRCBZ21} as the backbone, effectively leveraging the dual-path framework's ability to capture both long- and short-term dependencies in sequences, along with the powerful sequential modelling capability of Transformer. Secondly, our method benefits from the utilization of the DRL. Notably, our RSEN and GIDN are trained self-supervised. Despite the absence of speaker identity labels, our method still achieves SOTA performance. As illustrated in Table \ref{tab5}, when compared to X-SepFormer \cite{abs-2303-05023}, which also adopts Sepformer as the underlying architecture, our method attains better performance by incorporating DRL and AM-Transformer. Our approach showcases a significant decrease of $1.36$ in $r_{\text{scr}}$, indicating its efficacy in addressing the SC problem.

\begin{table}[tb]
    \small
    \centering
    \setlength\tabcolsep{2pt}
    \begin{tabular}{l|ccc|c}
        \toprule
        Method & SI-SNRi$\uparrow$ & SDRi$\uparrow$  & PESQ $\uparrow$ & $r_{\text{scr}}$ $\downarrow$\\ 
        \midrule
        TseNet \shortcite{XuRCL19} & $12.2$ & $12.6$ & $3.14$ & -\\
        SpEx \shortcite{XuRCL20} & $14.2$ & $14.6$  & $3.36$ &  $9.68$ \\
        SpEx+ \shortcite{GeXWCD020} & $15.7$ & $15.9$ & $3.49$ & $9.29$ \\
        DPRNN-Spe-IRA \shortcite{DengMSZZSW21} & $17.5$ & $17.7$ & $3.62$ & - \\
        SDR-TSE (Ours) &  $\textbf{19.9}$ &  $\textbf{20.2}$ & $\textbf{3.85}$ & $\textbf{6.89}$ \\
        \bottomrule
    \end{tabular}
    \caption{Performance comparison with SOTA methods on the WSJ0-2mix-extr dataset.}
    \label{tab4}
\end{table}

\begin{table}[tb]
    \centering
    \small
    \setlength\tabcolsep{2pt}    
    \begin{tabular}{l|ccc|c}
        \toprule
        Method & SI-SNRi$\uparrow$ & SDRi$\uparrow$ & PESQ $\uparrow$ & $r_{\text{scr}}$ $\downarrow$\\ 
        \midrule
        SpEx \shortcite{XuRCL20} & $15.8$ & $16.3$ & $3.14$  & $10.42$ \\
        SpEx+ \shortcite{GeXWCD020} & $16.9$ & $17.2$ & $3.45$ & $9.68$ \\
        DPRNN-Spe-IRA \shortcite{DengMSZZSW21} & $17.3$ & $17.6$ & $3.43$ & - \\
        X-SepFormer ($S_{sc}$) \shortcite{abs-2303-05023} & $19.1$ & $19.7$ & $3.75$ & $8.56$ \\
        X-SepFormer ($S_{wt}$) \shortcite{abs-2303-05023} & $18.8$ & $19.3$ & $3.74$ & $8.03$ \\
        SDR-TSE (Ours) & $\textbf{19.6}$ &  $\textbf{19.9}$  & $\textbf{3.82}$ & $\textbf{6.67}$  \\
        \bottomrule
    \end{tabular}
    \caption{Performance comparison with SOTA methods on the WSJ0-2mix dataset.}
    \label{tab5}
\end{table}

\section{Conclusion}

This paper introduces SDR-TSE, a novel approach to tackle the speaker confusion problem in TSE from the perspective of information disentanglement. Our self-supervised DRL policy disentangles the speaker identity information from the reference speech in two phases, providing effective guidance for TSE. Additionally, we propose the AM-Transformer, which integrates the AMLN to preserve the acoustic representation's information in the SEN and enhance its perception of speaker embeddings. Through extensive experiments, we showcase our meticulously designed method's strong information disentanglement capability and its exceptional performance in TSE.

\bibliography{aaai24}

\end{document}